\documentclass[12pt,letter]{article}
\pdfoutput=1
\usepackage{graphicx, epsfig, color,cite}
\usepackage{amsmath}
\usepackage{amssymb}
\usepackage{float}
\usepackage{subfig}
\usepackage{hyperref}

\newcommand{\MET}{E_T^{\rm miss}}

\def\eslt{\not\!\!\!{E_T}}
\def\to{\rightarrow}

\def\bi{\begin{itemize}}
\def\ei{\end{itemize}}

\def\tchi{\tilde\chi}

\def\tst{\tilde t}

\def\tg{\tilde g}

\def\alt{\lesssim}
\def\agt{\gtrsim}
\def\be{\begin{equation}}  
\def\ee{\end{equation}}  
\def\bea{\begin{eqnarray}}  
\def\eea{\end{eqnarray}}

\begin{document}
\begin{titlepage}
\begin{flushright}
OU-HEP-261015
\end{flushright}

\vspace{0.5cm}
\begin{center}
  {\Large \bf String phenomenology from the landscape}\\
\vspace{1.2cm} \renewcommand{\thefootnote}{\fnsymbol{footnote}}
{\large Howard Baer$^{1}$\footnote[1]{Email: baer@ou.edu },
Vernon Barger$^2$\footnote[2]{Email: barger@pheno.wisc.edu},
Shadman Salam$^3$\footnote[3]{Email: shadman.salam@ewubd.edu}
  and Kairui Zhang$^1$\footnote[4]{Email: kzhang25@ou.edu}
}\\ 
\vspace{1.2cm} \renewcommand{\thefootnote}{\arabic{footnote}}
{\it 
$^1$Homer L. Dodge Department of Physics and Astronomy,
University of Oklahoma, Norman, OK 73019, USA \\[3pt]
}
{\it 
$^2$Department of Physics,
University of Wisconsin, Madison, WI 53706 USA \\[3pt]
}
{\it
  $^3$Department of Mathematics and Data Science, East West University, A-2 Jahurul Islam Avenue, Aftabnagar, Dhaka 1212, Bangladesh} \\[3pt]
\end{center}

\vspace{0.5cm}
\begin{abstract}
\noindent
SUSY breaking trilinear terms, known as $A$-terms, play a key role
in supersymmetric phenomenology in that 1. large weak scale values help
boost the light Higgs boson mass $m_h\to 125$ GeV and 2. the same large
negative values {\it decrease} the top-squark contributions to the weak scale,
thus making the theory more natural. Their renormalization group
evolution also plays a key role in string landscape in that large values can
drive top-squark soft terms to tachyonic values leading to
charge-and-color breaking (CCB) minima in the scalar potential.
Adjacent to these CCB regions is where $m_h\to 125$ GeV and where natural
parameter space can be found. This is an example of {\it living dangerously}
on the landscape where the draw to large soft terms stops just short
of catastrophic minima, but where nature is natural with a Higgs mass near its measured value. 
In a final panoramic view, we compare the reach of HL-LHC for landscape SUSY via several prominent signatures in the higgsino vs. gaugino mass parameter space for the case of decoupled first/second generation scalars.
\end{abstract}
\end{titlepage}

\section{Introduction}
\label{sec:intro}

Gordy Kane {\it et al.}\cite{Kane:2014bau} have emphasized, correctly,
a bidirectional (nutcracker) approach to string phenomenology:
1. balancing a bottom-up approach with the $N=1$ supersymmetry (SUSY) augmented
Standard Model as a target Quantum Field Theory (QFT)  with
2. a top-down approach starting with compactified string theory.
This process was begun by Candelas {\it et al.}\cite{Candelas:1985en}
who identified heterotic string theory compactified to $M_4\times X_6$
on a Calabi-Yau 3-fold
with a special holonomy group $SU(3)$ admitting exactly one covariantly constant
spinor leading to a remnant $N=1$ spacetime SUSY on $M_4$.
Since then, other compactification possibilities have been developed,
leading to $N=1$ spacetime SUSY in $4-$dimensions, as shown in
Table \ref{tab:susy}.
\begin{table}[!htb]
\renewcommand{\arraystretch}{1.2}
\begin{center}
\begin{tabular}{c|c}
string & manifold \\
\hline
heterotic & $CY_3$ with special holonomy $SU(3)$ \\
$IIA/IIB$  & $CY_3+$ orientifold projection \\
$I$   & $CY_3+$ orientifold \\
$M-$theory & $7-$manifold with $G_2$ holonomy \\
$F-$theory & elliptically fibered $CY_4$ \\
\hline
\end{tabular}
\caption{Various string theory compactifications leading to $N=1$
  spacetime SUSY on $M_4$.
}
\label{tab:susy}
\end{center}
\end{table}
It has become appreciated in the 21st century that the number of viable
(flux) compactification possibilities\cite{Douglas:2006es} range from at
least $10^{500}$\cite{Ashok:2003gk} to possibly many, many more, each
giving a different set of $4-d$ laws of physics.
This situation makes possible Weinberg's anthropic solution to the cosmological constant problem\cite{Weinberg:1987dv} but also makes identification of the correct compactification manifold
used by nature a daunting if not impossible task. A way forward proposed by Douglas\cite{Douglas:2004qg} is for {\it statistical} explorations of the string landscape, a situation we investigate here in the context of soft SUSY breaking.

\subsection{Lessons from the CERN LHC}

\begin{figure}[htb!]
\centering
    {\includegraphics[height=0.4\textheight]{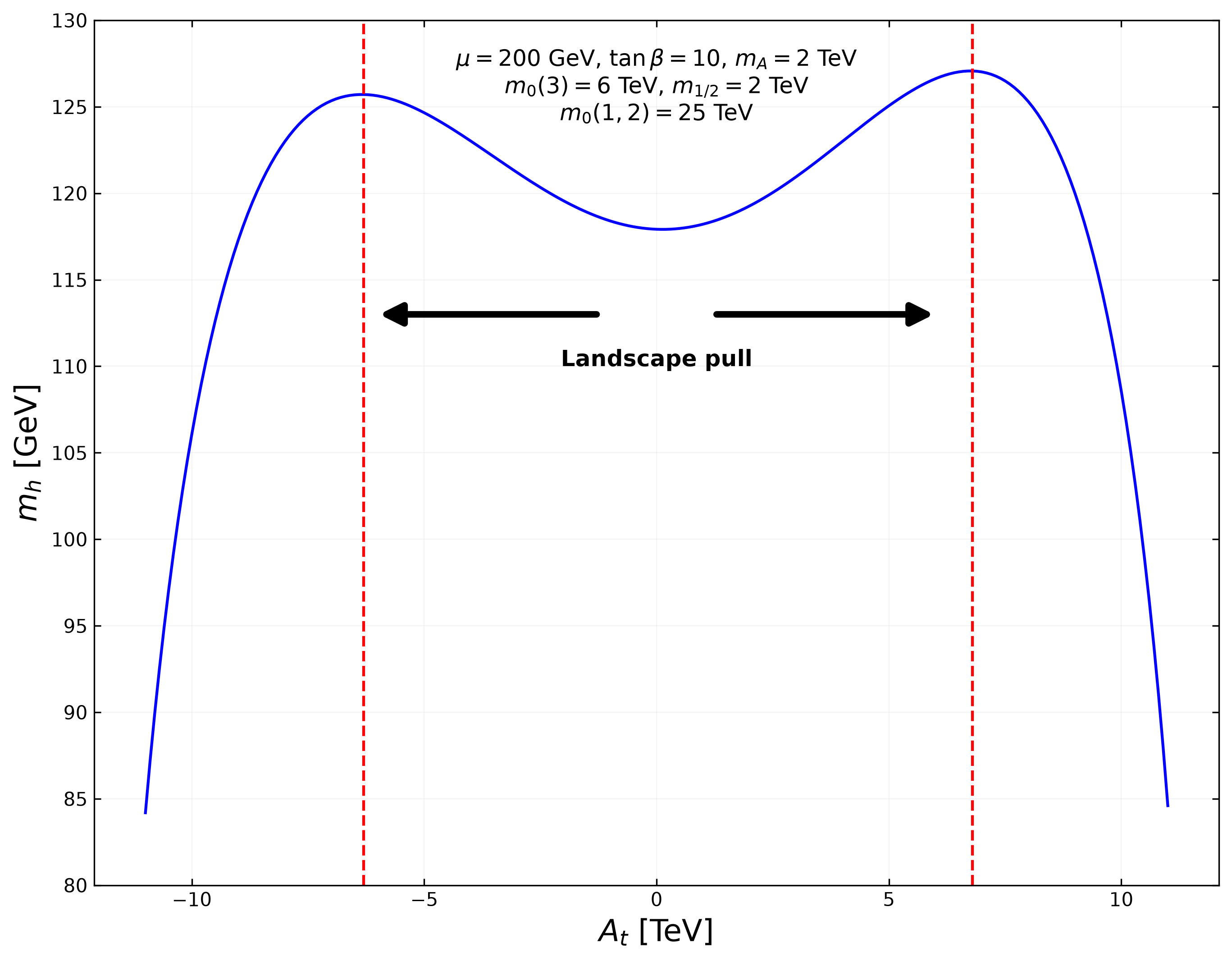}}
    \caption{Plot of $m_h$ vs. $A_t$ for the natural SUSY MSSM benchmark point.
      Calculation of SUSY spectrum from Isajet 7.92\cite{Paige:2003mg}
      and calculation of $m_h$ from FeynHiggs\cite{Bahl:2018qog}.
      \label{fig:mh}}
\end{figure}

What have we learned from the CERN LHC over the past fifteen years?
That there exists a very Standard Model (SM)-like Higgs boson with
mass $m_h\simeq 125$ GeV\cite{ATLAS:2012yve,CMS:2012qbp}, with as yet no
sign of the expected
supersymmetric (SUSY) matter\cite{Baer:2006rs,ATLAS:2024lda,Sekmen:2025bxv,Jeanty:2026etw,Baer:2025zqt}.
The Higgs mass is not inconsistent with weak scale SUSY, since
in the MSSM $m_h$ is bounded by $m_h\alt 135$ GeV\cite{Carena:2002es},
and is given (including dominant 1-loop corrections) by
\be
m_h^2=m_Z^2\cos^2 2\beta +\frac{3 g^2 m_t^4}{8\pi^2m_W^2}
\left(\log \left(\frac{m_{SUSY}^2}{m_t^2}\right)+ \frac{x_t^2}{m_{SUSY}^2}\left(1-\frac{x_t^2}{12 m_{SUSY}^2}\right)\right)
\label{eq:mh}
\ee
where $m_{SUSY}^2\simeq m_{\tst_1}m_{\tst_2}$ optimizes the scale choice and
$x_t=A_t-\mu\cot\beta$.
The value of $m_h^2$ is maximal for $x_t=\pm\sqrt{6}m_{SUSY}$ 
(the maximal mixing scenario\cite{Carena:2002es})
which then requires rather large weak scale trilinear soft terms ($A$-terms).
This is illustrated in Fig. \ref{fig:mh} where we plot the value of $m_h$
(as computed with FeynHiggs\cite{Bahl:2018qog}) for a natural SUSY
benchmark point generated with Isajet 7.92\cite{Paige:2003mg}
in the NUHM3 model\cite{Ellis:2002wv} with parameters as shown in the Figure.
For small $A_t\sim 0$, then $m_h$ is suppressed to values $\alt 120$ GeV.
One way to increase the value of $m_h$ is by jacking up the top-squark masses
to such high values that the loop contributions $\Sigma_u^u(\tst_{1,2})$ to the weak scale\cite{Baer:2012cf} become huge, and the model becomes EW-{\it unnatural}.
Alternatively, if $|A_t|$ is large, then $m_h\to 125$ GeV
(and large $A_t$ also decreases the $\Sigma_u^u(\tst_{1,2})$ values so the
model becomes more natural\cite{Baer:2012up}.

Some popular models such as gauge mediation\cite{Dine:1995ag,Giudice:1998bp}
(GMSB), anomaly-mediation\cite{Giudice:1998xp,Randall:1998uk} (AMSB)
and gaugino mediation\cite{Chacko:1999mi},
expect tiny $A$-terms and so require large, unnatural values of top-squark
soft masses to accommodate the measured value of $m_h$\cite{Baer:2014ica}.
In contrast, in gravity-mediation\cite{Cremmer:1982en,Nilles:1983ge},
large $A$-terms of order $A\sim m_{3/2}\sim m_{weak}$ are expected.
In this sense, LHC observations seem to select out gravity-mediated SUSY
breaking (SUGRA) as the low energy effective field theory (EFT) with
broken SUSY.

\subsection{Soft term expectations from SUGRA}

In SUGRA models, one assumes a hidden sector which serves as an arena for
SUSY breaking\cite{Nilles:2001ap}, and a visible sector containing
(at least) the usual MSSM fields.
All are contained in the SUGRA Lagrangian\cite{Cremmer:1982en},
which is expected to be the low energy effective field theory (EFT) of
string compactification to four dimensions\cite{Witten:1985xb}.
Under hidden sector SUSY breaking ({\it e.g.} via gaugino condensation\cite{Ferrara:1982qs,Dine:1985rz}),
  the weak scale soft SUSY breaking terms
emerge from Planck-suppressed SUGRA operators.

\subsubsection{Scalar soft SUSY breaking terms}

Scalar soft breaking masses arise from 
\be
\int d^4\theta c_{ij}\frac{X^\dagger X Q_i^\dagger Q_j}{m_P^2}\to
c_{ij}\frac{F_X^\dagger F_X}{m_P^2}\phi_i^*\phi_j
\label{eq:mscalar}
\ee
where the $X$ fields are hidden sector SUSY breaking fields and the
$Q$ are visible sector chiral superfields and the $\theta$s are
superspace coordinates.
When the auxiliary $X$ fields gain a SUSY breaking vev $F_X$, then
scalar soft squared masses $m_{ij}^2\sim c_{ij}\frac{F_X^\dagger F_X}{m_P^2}$
arise and if $F_X\sim (10^{11}$ GeV)$^2$, then weak scale
soft scalar mass terms will ensue.

In the above, we leave the scalar field indices $i,j$ as arbitrary
so long as $Q_i^\dagger Q_j$ is invariant under the assumed
(gauge and possibly other) symmetries; this emphasizes that
{\it nonuniversal scalar masses}\cite{Soni:1983rm,Kaplunovsky:1993rd,Brignole:1993dj}
are the {\it rule and not the exception} in SUGRA, and
they also lead to the well-known SUSY flavor problem\cite{Hall:1985dx}.
Thus, models like CMSSM/mSUGRA\cite{Arnowitt:1993qp,Kane:1993td}
with an assumed ad-hoc scalar mass universality are not likely a realistic
expression of gravity-mediation. 
Also, since the scalar masses arise from
$F_X^\dagger F_X$, they are not protected by any hidden sector symmetries
which might enforce small (weak scale) values. One form of universality is
highly motivated: intra-generational scalar mass degeneracy since the matter of
each generation lives in the 16-dimensional spinor rep of $SO(10)$. 
Thus, in SUGRA theories, we expect the soft scalar masses at the GUT unification 
scale $Q\sim m_{GUT}$ to be
$m_{H_u}\ne m_{H_d}\ne m_0(1)\ne m_0(2)\ne m_0(3)$ which is accommodated 
by the so-called four-extra-parameter non-universal Higgs model 
NUHM4\cite{Ellis:2002wv,Baer:2005bu}.

\subsubsection{Gaugino masses}

Gaugino masses in SUGRA originate from
\be
\int d^2\theta \frac{Y W^\alpha W_\alpha}{m_P}\to \frac{F_Y}{m_P}\lambda\lambda
\label{eq:mino}
\ee
leading to weak scale gaugino masses when the hidden sector field $Y$
attains a SUSY breaking vev $F_Y$.
It is important to notice here that the SUSY breaking field $Y$ must be a
gauge singlet and further that
generation of gaugino masses leads to a broken $R$-symmetry in
the Lagrangian since the gauginos carry $R$-charge $+1$.
Indeed, non-singlet hidden sector SUSY breaking was one of the main
motivations for introducing the anomaly-mediated form for gaugino masses\cite{Giudice:1998xp}.

\subsubsection{$A$-terms}

Trilinear terms arise as
\be
\int d^2\theta \frac{Y H_uQ U^c}{m_P}\to \frac{F_Y}{m_P} H_u\tilde{Q}\tilde{u}_R
\label{eq:Aterms}
\ee
{\it etc.} so that trilinear soft terms $a_{ijk}\sim m_{weak}$ also develop.
As for gaugino masses, the field $Y$ must be a hidden sector singlet
in order to develop large $A$-terms
(as needed by the rather large measured value of $m_h\simeq 125$ GeV).
Like gaugino masses, if the starting Lagrangian is symmetric under a
continuous $R$-symmetry
($R(H_uQ U^c)=+2$ to balance against $R(d^2\theta )=-2$), then $A$-terms would
also be forbidden under $F$-term breaking by the Nelson-Seiberg theorem\cite{Nelson:1993nf}.
The development of metastable SUSY breaking models\cite{Intriligator:2006dd,Dine:2010cv} sidesteps this issue
and allows for gaugino masses and $A$-terms of order $m_{soft}\sim m_{3/2}$.

\subsubsection{$B$-terms}

Bilinear soft terms come from the operator
\be
   \int d^4\theta X^\dagger X H_u H_d/m_P^2\to B_\mu\sim (F_X/m_P)^2 H_u H_d
   \ee
and are also of order the weak scale.

\subsection{$\mu$ term}

The SUSY conserving $\mu$ term would be expected to arise with $\mu\sim m_P$
whereas phenomenology requires $\mu\sim m_{weak}$.
This is the SUSY $\mu$ problem\cite{Bae:2019dgg}.
First, $\mu$ must be forbidden
by some mechanism/symmetry, and then regenerated at the weak scale.
One neglects the SUSY $\mu$ problem at one's own peril.
Here we expect the $\mu$ term is initially forbidden by a remnant discrete
$R$-symmetry $\mathbb{Z}_n^R$ arising from string compactification\cite{Lee:2011dya,Nilles:2017heg}.
The $\mu$-less and unbroken MSSM then contains an accidental global
$U(1)_{PQ}$ symmetry which can be used for the SUSY DFSZ solution to the
strong CP problem\cite{Baer:2026jeb,Baer:2025srs,Baer:2025oid}.

The $\mu$ term may re-emerge with $\mu\sim m_{soft}$
via a Kim-Nilles\cite{Kim:1983dt} operator
\be
W\ni \lambda_\mu S^2H_uH_d/m_P
\ee
where $S$ develops a vev $\langle S\rangle\sim\sqrt{m_{weak}m_P}\sim 10^{11}$ GeV
as a consequence of SUSY breaking.
By Goldstone's theorem, a pseudo-Goldstone boson the axion emerges since the
global $U(1)_{PQ}$ is also broken. For strong enough $\mathbb{Z}_n^R$
({\it e.g.} $n=24$), then the axion quality problem is also solved\cite{Lee:2011dya,Baer:2018avn,Bhattiprolu:2021rrj}.

\section{Statistical selection of soft terms from the string landscape}
\label{sec:soft}

A third string revolution occurred when Bousso and Polchinski\cite{Bousso:2000xa} found that
string flux compactifications provided a setting for
Weinberg's anthropic solution\cite{Weinberg:1987dv} to the cosmological
constant problem. Each distinct compactification possibility gives rise to
different $4-d$ laws of physics depending on the details of the compact space, of which there are an enormous number of possibilities ($10^{500}$ possibilities
is a widely quoted number\cite{Ashok:2003gk},
but the actual value may be much larger\cite{Douglas:2006es}).

\subsection{Distribution of soft terms}

In vacua restricted to those containing the MSSM as the LE-EFT,\footnote{
  It is conjectured that only compactifications including a remnant
$N=1$ spacetime supersymmetry  result in stable vacua\cite{Acharya:2019mcu}.}
it is expected that the soft SUSY breaking terms will scan along with the CC.
Rather general arguments from Ref's \cite{Douglas:2004qg,Susskind:2004uv,Arkani-Hamed:2005zuc} imply a power-law draw to large soft terms on the string landscape.
In the simplest case of SUSY breaking via a single (complex values) hidden sector $F_X$-term-- since nothing in string theory favors
one value of $\langle F_X\rangle$ over another-- then the values of $F_X$
should be randomly distributed as a complex number.
This then leads to a linear distribution in the SUSY breaking scale $m_{hidden}$
since $m_{hidden}^4\sim F_X^\dagger F_X$ (a circle in the complex $F_X$ plane).
For cases with more hidden sector fields participating in SUSY breaking, then
instead one expects the distribution of soft terms $f_{SUSY}$ to be
\be
f_{SUSY}\sim m_{soft}^{2 n_F+n_D-1}
\label{eq:fSUSY}
\ee
where $n_F$ is the number of breaking fields and $n_D$ is the number of $D$-breaking fields participating in the overall SUSY breaking.
Due to their different dependencies on the hidden sector moduli fields, the
different soft terms are expected to scan independently of each
other\cite{Baer:2020vad}.

\subsection{Large soft terms tempered by atomic principle}

Of course, if large soft terms are statistically favored over small ones,
one might expect all soft terms up around the Planck scale.
However, we must insist on the existence of nuclei/atoms which can give
rise to complexity, and this means requiring
\begin{enumerate}
\item appropriate electroweak symmetry breaking (EWSB) and
  \item a derived value of the weak scale
in each pocket universe (PU) (within the greater multiverse) which lies within the ABDS window of
allowed values\cite{Agrawal:1997gf}:
\be
m_Z^{OU}/2\alt m_Z^{PU}\alt 4m_Z^{OU} .
\label{eq:abds}
\ee
\end{enumerate}
Here, $m_Z^{PU}$ is the derived pocket universe Z-mass, while $m_Z^{OU}=91.2$ GeV
is the value of $m_Z$ in our universe.

The first of these conditions means that EWSB is properly broken via vevs
of the two Higgs doublet fields $v_u$ and $v_d$, where $\tan\beta\equiv v_u/v_d$, and where $m_{H_u}^2$ is driven radiatively to negative values causing the
Higgs potential to develop an appropriate minimum\cite{Ibanez:1982fr,Alvarez-Gaume:1983drc}.
It also means that we would veto models where charge-or-color breaking (CCB)
minima develop, which is signaled by one of the charged or colored scalar
fields (usually $m_{U_3}^2$) being driven to negative values.

The second condition arises from ABDS\cite{Agrawal:1997gf} in that too large
a value of the weak scale ($\agt 4m_Z^{OU}$) results in no stable nuclei and
too small a value results in nuclei of all neutrons with protons being unstable.
Only models wherein the derived weak scale $m_Z^{PU}$ is within the ABDS window
are then acceptable.

In the MSSM, the derived value of $m_Z$ is given by
\be
m_Z^2/2 =\frac{m_{H_d}^2+\Sigma_d^d-(m_{H_u}^2+\Sigma_u^u)\tan^2\beta}{\tan^2\beta -1}-\mu^2
\label{eq:mzs}
\ee
where the $\Sigma_u^u$ and $\Sigma_d^d$ contain an assortment of loop
corrections computed in the Coleman-Weinberg formalism\cite{Coleman:1973jx}
and listed in Ref's \cite{Baer:2012cf} and \cite{Baer:2021tta}.
Ordinarily in SUSY spectrum code programs,
either the value of $\mu^2$ or $m_{H_u}^2$ is tuned in order to guarantee
that $m_Z=m_Z^{OU}=91.2$ GeV. Without finetuning, then typically
$m_Z^2/2$ is (nearly) given by the maximal value on the right-hand-side (RHS)
of Eq. \ref{eq:mzs}. The maximal value is actually used to compute the
{\it electroweak} finetuning measure $\Delta_{EW}$\footnote{The Isasugra
  spectra code\cite{Paige:2003mg}
  (which we use here for all SUSY spectra computations)
  contains a computation for $\Delta_{EW}$ and so does the
publicly available code \cite{Baer:2023cvi} \it{DEW4SLHA}.}
In this case, the derived value of the weak scale in different string compactifications is then given by
\be
m_Z^{PU}\equiv m_Z^{OU}\sqrt{\Delta_{EW}/2} .
\ee

\subsection{Natural SUSY emergent from landscape}

In this paper, we will work within the framework of the 4-extra-parameter
non-universal Higgs model (NUHM4), which we expect is the form the
MSSM should take under gravity-mediation.
The NUHM4 parameter space is given by
\be
m_0(1),\ m_0(2),\ m_0(3),\ m_{H_u},\ m_{H_d},\ m_{1/2},\ A_0,\ B\ \ \ (NUHM4)
\ee
where $B$ is usually traded for the ratio of Higgs vevs $\tan\beta$ via
the electroweak minimization conditions\cite{Baer:2006rs}.
The unification of matter within the generations is motivated by the fact
that all matter superfields of each generation fill out the 16-d spinor
rep of $SO(10)$.
Split representations (matter superfields versus Higgs and gauge superfields)
occur naturally in {\it local} SUSY GUTs that arise in
string compactifications, where the locus of various fields on the compact
space determines which symmetries they obey\cite{Buchmuller:2005sh,Nilles:2009yd}.
Gaugino mass unification is motivated if the various gauge superfields couple
directly to the same SUSY breaking field $F_Y$\cite{Brignole:1993dj}.
The various soft terms are expected to scan according to the power-law
distribution given in Eq. \ref{eq:fSUSY}.

The various soft terms are then selected to be as large as possible, subject to
a livable vacuum state.
\begin{itemize}
\item $m_0(1)$ and $m_0(2)$: Since first/second generation matter have tiny
  Yukawa couplings, they contribute through these at a small level
  to the weak scale. Their upper bounds usually occur at the
  $m_0(1,2)\alt 20-40$ TeV-level via 2-loop RG effects which can drive
  third generation soft terms to tachyonic values\cite{Arkani-Hamed:1997opn,Baer:2010ny,Baer:2024hpl}.
\item $m_0(3)$: If this soft term gets too large, then the
  $\Sigma_u^u(\tst_{1,2})$ contributions drive the weak scale to values beyond
  the ABDS window.
\item $m_{H_u}$: For small high scale values $m_{H_u}^2(\Lambda )$,
  then $m_{H_u}^2$ is driven to deep negative  unnatural values.
  The larger $m_{H_u}^2(\Lambda )$ is, the more natural is its weak scale value
  (radiatively driven naturalness or RNS) until it gets too big, and no longer is driven negative at the weak scale, so EWSB no longer occurs.
\item $m_{H_d}^2$: For large, decoupled values, then $m_{H_d}\sim m_{A,H,H^\pm}$.
  If $m_{H_d}/\tan\beta$ gets too large, it pushes $m_Z^{PU}$ beyond the
  ABDS window by Eq. \ref{eq:mzs}.
\item $m_{1/2}$: If $m_{1/2}$ gets too large, the gluino mass becomes so large it pushes the $\Sigma_u^u(\tst_{1,2})$ to too large values. It also
  contributes to $m_{weak}$ via 2-loop corrections\cite{Dedes:2002dy}.
\item $A$-terms: (The topic of this paper.) If $A_t$ becomes large, then
  cancellations in the $\Sigma_u^u (\tst_{1,2})$ occur making the spectra more natural. If $A_t$ becomes too large, it contributes to driving $m_{U_3}^2<0$
  leading to CCB minima in the scalar potential.
\item $B$-term: For $B$ too big, then $\tan\beta \to 1$ and
  terms like $m_{H_d}/\tan\beta$ in Eq. \ref{eq:mzs} become too large
  while $m_h$ is suppressed to $\ll 125$ GeV.
  \end{itemize}

In this paper, we investigate the consequences of $A$ and $B$ term running for string landscape spectra.

\section{$A$-terms}
\label{sec:Aterms}

During 20th and early 21st century SUSY phenomenology, the trilinear soft terms
seemed rather inconsequential, and indeed were set to zero in many
plots.
Since the Higgs discovery, they have assumed a crucial role, mainly because
the MSSM Higgs mass is maximized for large $A_t(weak)$ values,
as shown in Eq. \ref{eq:mh} and Fig. \ref{fig:mh} and Fig. 1 of Ref. \cite{Baer:2024fgd}.
Shortly after the Higgs discovery, it was commonly lamented that SUSY was
unnatural due to the rather large value of $m_h\sim 125$ GeV.
But now it may be appreciated that large $m_h$ tends to select out
gravity-mediated SUSY breaking, where large $A$-terms are expected,
and where abnormal conditions need to be invoked to suppress them in the
SUGRA framework.
It was also early-on lamented that large $A$-terms gave rise to unnatural
SUSY under the naturalness measure
$\Delta_{HS}$\cite{Brust:2011tb,Papucci:2011wy,Hall:2011aa}.
However, the opposite is true under the measure $\Delta_{EW}$\cite{Baer:2012up},where top-squark contributions to the weak scale are given at 1-loop by
\be
\Sigma_u^u(\tst_{1,2})=\frac{3}{16\pi^2}F(m_{\tst_{1,2}}^2)
\left[ f_t^2-g_Z^2\mp\frac{f_t^2A_t^2-8g_Z^2(\frac{1}{4}-\frac{2}{3}x_W)\Delta_t}{m_{\tst_2}^2-m_{\tst_1}^2}\right]
\label{eq:Sigmauu}
  \ee
where $\Delta_t=(m_{\tst_L}^2-m_{\tst_R}^2)/2+m_Z^2\cos 2\beta (\frac{1}{4}-\frac{2}{3}x_W)$
and $x_W\equiv\sin^2\theta_W$ and $F(m^2 )=m^2\left(\log\frac{m^2}{Q^2}-1\right)$ with an optimized scale choice $Q^2=m_{\tst_1} m_{\tst_2}$.
  Large $A_t$ leads to large cancellations in both of the
  $\Sigma_u^u (\tst_{1,2})$ terms leading to increased naturalness.

The RGEs for (1-loop) $A_{t,b,\tau }$ term evolution are given by 
\bea
\frac{dA_t}{dt} &=& \frac{2}{16\pi^2}\left( -\sum_i c_ig_i^2M_i+6 f_t^2A_t+f_b^2 A_b\right)\\
\frac{dA_b}{dt} &=& \frac{2}{16\pi^2}\left( -\sum_i c_i^\prime g_i^2M_i+
6 f_b^2A_b+f_t^2 A_t+f_\tau^2A_\tau \right)\\
\frac{dA_\tau}{dt} &=& \frac{2}{16\pi^2}\left( -\sum_i c_i^{\prime\prime}g_i^2M_i+
3f_b^2 A_b+4f_\tau^2 A_\tau\right)
\label{eq:dAdt}
\eea
where $t=\log (Q)$ and $f_{t,b}$ are top/bottom Yukawa couplings, $A_{t,b,\tau}$
are trilinear ($A$-terms), the $M_i$ are gaugino masses, the $g_i$ are gauge
couplings and the index $i$ runs over the SM gauge groups.
The $c_i$s are $(13/15,3,16/3)$, $c_i^\prime$ are $(7/15,3,16/3)$ and
$c_i^{\prime\prime}$ are $(9/5,3,0)$.

\begin{figure}[htb!]
\centering
    {\includegraphics[height=0.4\textheight]{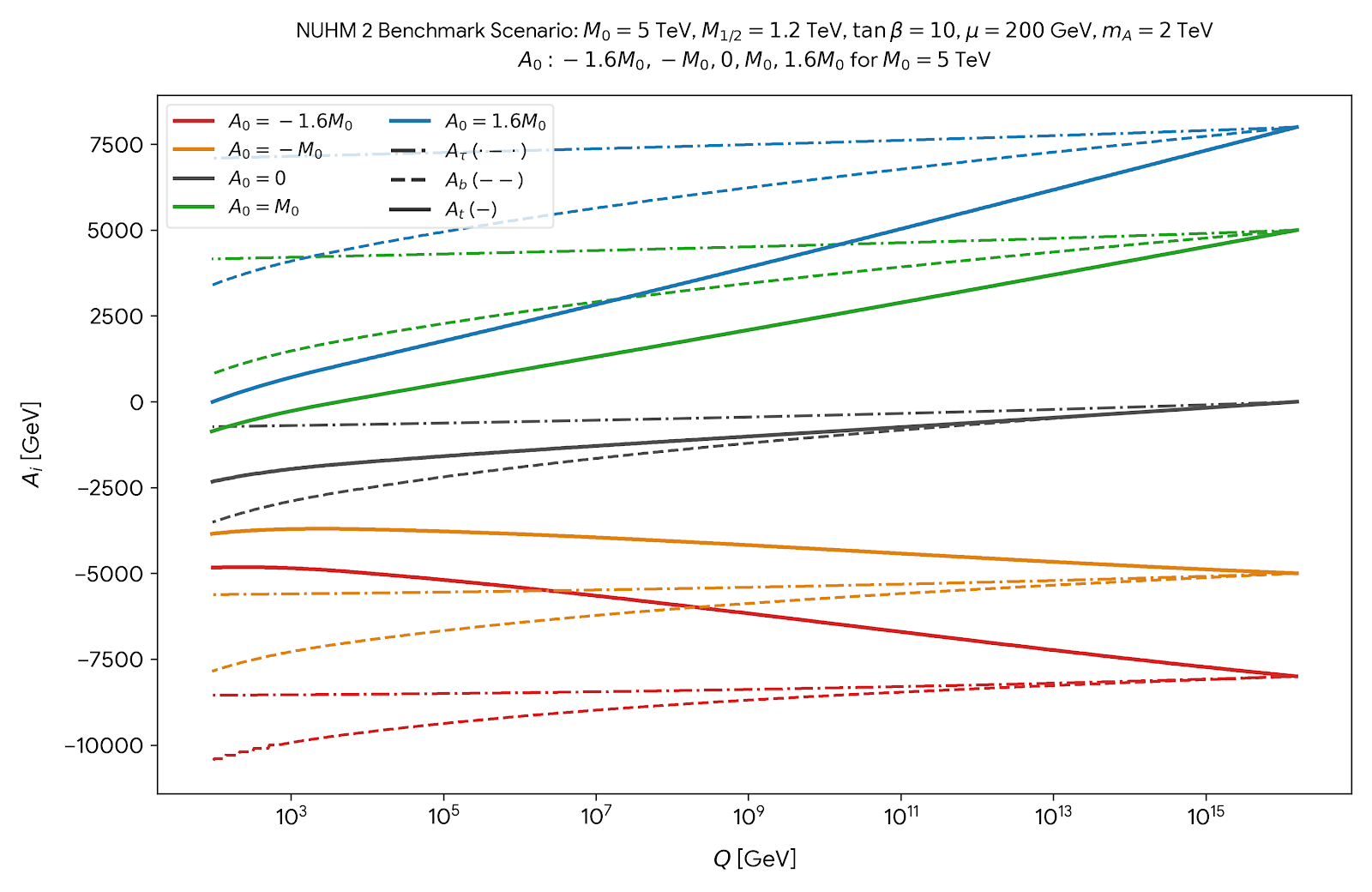}}
    \caption{Plot of evolution of third generation trilinear
      soft terms for five different choices of initial conditions.
      \label{fig:Aterms}}
\end{figure}

In Fig. \ref{fig:Aterms}, we plot the evolution of $A_{t,b,\tau}$
versus energy scale $Q$ from the unification scale $Q=m_{GUT}$ to the
weak scale, for a NUHM2 model benchmark point with $m_0=5$ TeV,
$m_{1/2}=1.2$ TeV, $\tan\beta =10$, $\mu =200$ GeV and $m_A=2$ TeV.
We take $A_0=1.6 m_0$ (blue), $A_0=m_0$ (green), $A_0=0$ (black),
$A_0=-m_0$ (orange) and $A_0=-1.6 m_0$ (red). Solid curves denote $A_t$,
dashed are $A_b$ while dot-dashed are $A_\tau$.
From the plot, we see that the $A_{\tau}$ terms hardly evolve; this is because
for $\tan\beta =10$, the $f_b$ and $f_\tau$ Yukawa couplings are small and the
$SU(3)_C$ gauge term doesn't contribute, so the slope of the curves is small.
There is much more change involved in the $A_t$ evolution where for
$A_0=+1.6 m_0$, the value of $A_t$ evolves from large GUT scale values to
small weak scale values: this leads to lower $m_h\alt 123$ GeV with increased
finetuning. The largest (in absolute value) of the $A_t(weak) $ values occurs
for $A_0\sim -1.6m_0$, where $A_t$ evolves from $-8$ TeV at $Q=m_{GUT}$
to $\sim -5$ TeV at $Q\sim m_{weak}$.
Such a large (negative) value of $A_t(weak)$ helps push $m_h\to 125$ GeV
while producing increased naturalness via cancellations in Eq. \ref{eq:Sigmauu}.
The evolution of $A_b$ is typically intermediate between the evolution
of $A_t$ and $A_\tau$.

One of the biggest effects of large $A_t$ terms is their impact on
generating CCB vacua.
These vacua arise from the running of the third generation stop soft term
$m_{U_3}^2$ via the RGE
\be
\frac{dm_{U_3}^2}{dt}=\frac{2}{16\pi^2}\left(-\frac{16}{15}g_1^2M_1^2-
\frac{16}{3}g_3^2M_3^2-\frac{2}{5}g_1^2 S+2f_t^2X_t\right)
\label{eq:dmU3dt}
\ee
where $S=m_{H_u}^2-m_{H_d}^2+Tr[{\bf m}_Q^2-{\bf m}_L^2-2{\bf m}_U^2+{\bf m}_D^2+{\bf m}_E^2]$ and $X_t=m_{Q_3}^2+m_{U_3}^2+m_{H_u}^2+A_t^2$.
The last term in Eq. \ref{eq:dmU3dt} is amplified by the large top Yukawa
coupling and the factor 2.
It can be further amplified by large $A_t^2$ which enters the expression
for $X_t$.
When this amplification occurs, then $m_{U_3}^2$ may be driven to negative
values at the weak scale thus generating CCB vacua.
\begin{figure}[htb!]
\centering
    {\includegraphics[height=0.4\textheight]{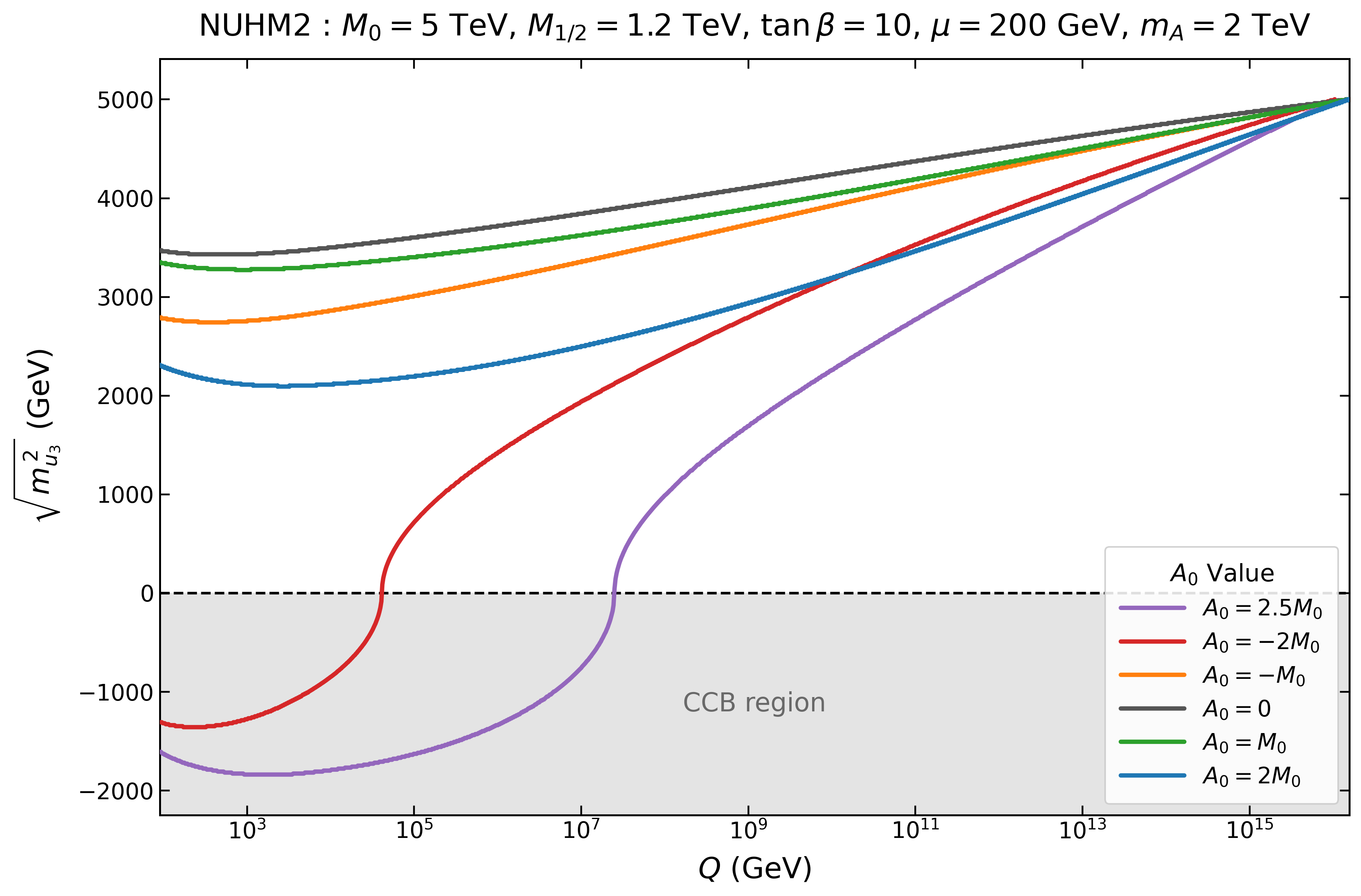}}
    \caption{Plot of evolution of $m_{U_3}^2$ soft term
       for six different choices of initial conditions.
      \label{fig:mU3}}
\end{figure}

This is shown in Fig. \ref{fig:mU3}, where we plot the running of $m_{U_3}^2$
vs. $Q$ for the same NUHM2 benchmark as in Fig. \ref{fig:Aterms} but now
for various values of $A_0/m_0$. For $A_0=0$, then $m_{U_3}^2$ runs to its
largest value shown at the weak scale. But for $A_0=\pm m_0$, then the downward
running from $Q=m_{GUT}$ to $Q=m_{weak}$ is enhanced, yielding lower, more natural
values of top-squark soft terms. For the red and purple curves, with
$A_0=-2m_0$ and $A_0=+2.5 m_0$ respectively, the value of $m_{U_3}^2$ is driven
negative before $Q=m_{weak}$ resulting in the generation of vacua with
charge and color breaking.
Such CCB vacua must be ruled out as viable anthropic vacuum states.

\section{$B$-term evolution}
\label{sec:B}

The bilinear soft term $V\ni bH_uH_d+h.c.$ is often taken as $b=B\mu$,
with the $B$ term evolution given by
\be
\frac{dB}{dt} = \frac{2}{16\pi^2}\left( -\frac{3}{5}g_1^2M_1-3g_2^2M_2+3f_t^2A_t+
3f_b^2A_b+f_\tau^2 A_\tau\right) .
\label{eq:dBdt}
\ee
The $B$-term evolution is usually ignored since the EW minimization conditions (at tree level) give
\be
B=\frac{(m_{H_u}^2+m_{H_d}^2+2\mu^2)\sin2\beta}{2\mu}
\label{eq:B}
\ee
and so allow for the replacement of $B(weak)$ by $\tan\beta$;
and then, $B$ is allowed to evolve to whatever value of $B(m_{GUT})$ is needed.

\begin{figure}[htb!]
\centering
    {\includegraphics[height=0.4\textheight]{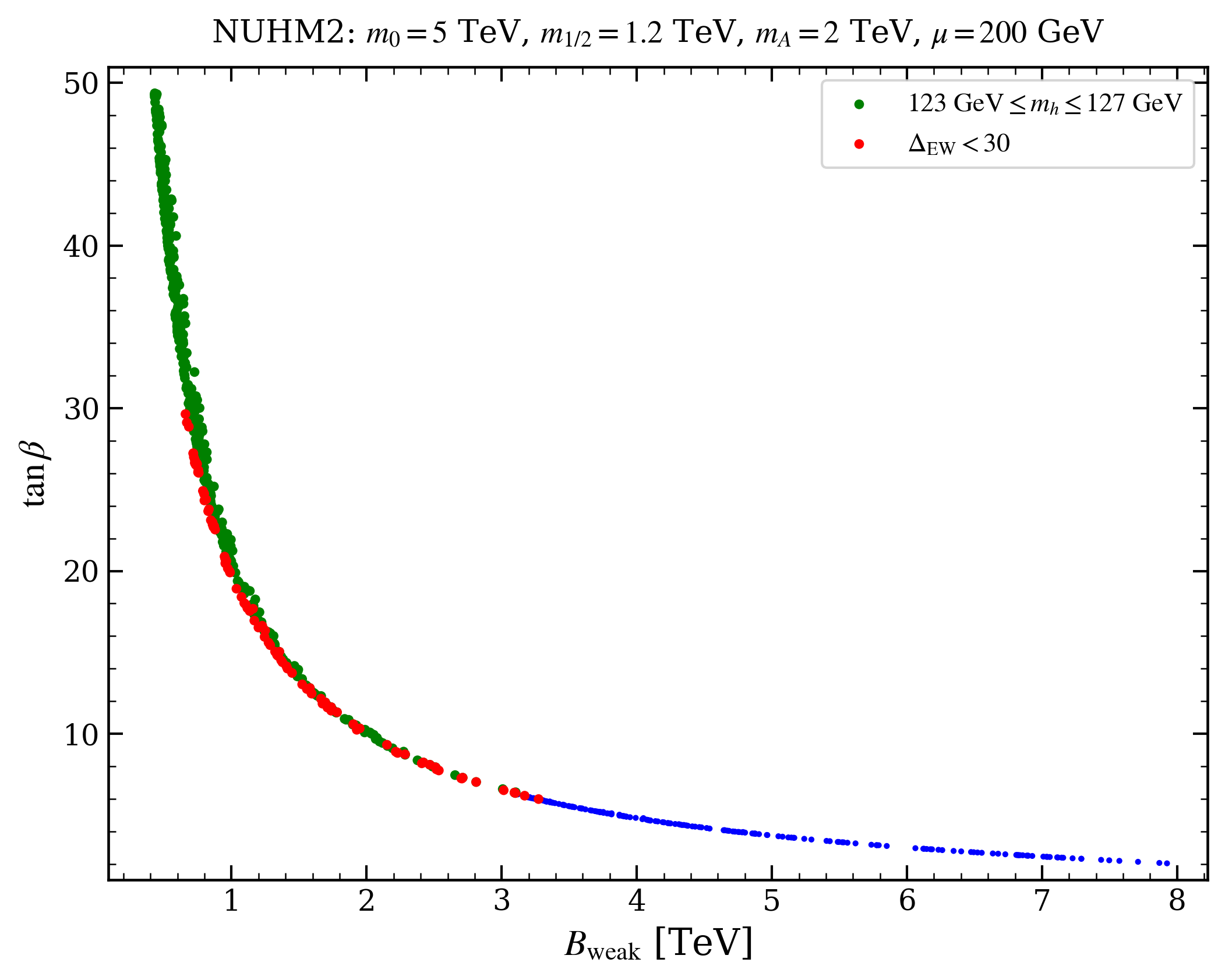}}
    \caption{Plot of $B(weak)$ vs. $\tan\beta$ in the NUHM2 model with
      $m_0(1,2,3)=5$ TeV, $m_{1/2}=1.2$ TeV and $m_{H_d}=2$ TeV,
      with $m_{H_u}=\mu =200$ GeV. The green line has $m_h: 123-127$ GeV
      while the blue line has $m_h<123$ GeV. The red region has $\Delta_{EW}<30$.
            \label{fig:BvTanb}}
\end{figure}

The value of $\tan\beta$ vs. $B(weak)$ for our usual BM point is shown in
Fig. \ref{fig:BvTanb}.
From the Figure, we see that for very large values of $B(weak)$,
then rather small values of $\tan\beta\sim 1$ are found.
For small $\tan\beta$, then the $m_{H_d}/\tan\beta$ contribution to the weak
scale in Eq. \ref{eq:mzs} may get large, thus pushing the value of $m_Z^{PU}$
well beyond the ABDS window. 
Thus, large $B(weak)$ and large $m_{H_d}(weak)$ have trouble coexisting with 
each other. For very small $B(weak)$, then $\tan\beta \agt 30$ and the $b$- and $\tau$-Yukawa couplings become large, leading to large 3rd generation contributions to $\Delta_{EW}$. The intermediate region $\tan\beta\sim 6-30$
allows for naturalness ($\Delta_{EW}\alt 30$) as shown by the red-shaded region.

\begin{figure}[htb!]
\centering
    {\includegraphics[height=0.4\textheight]{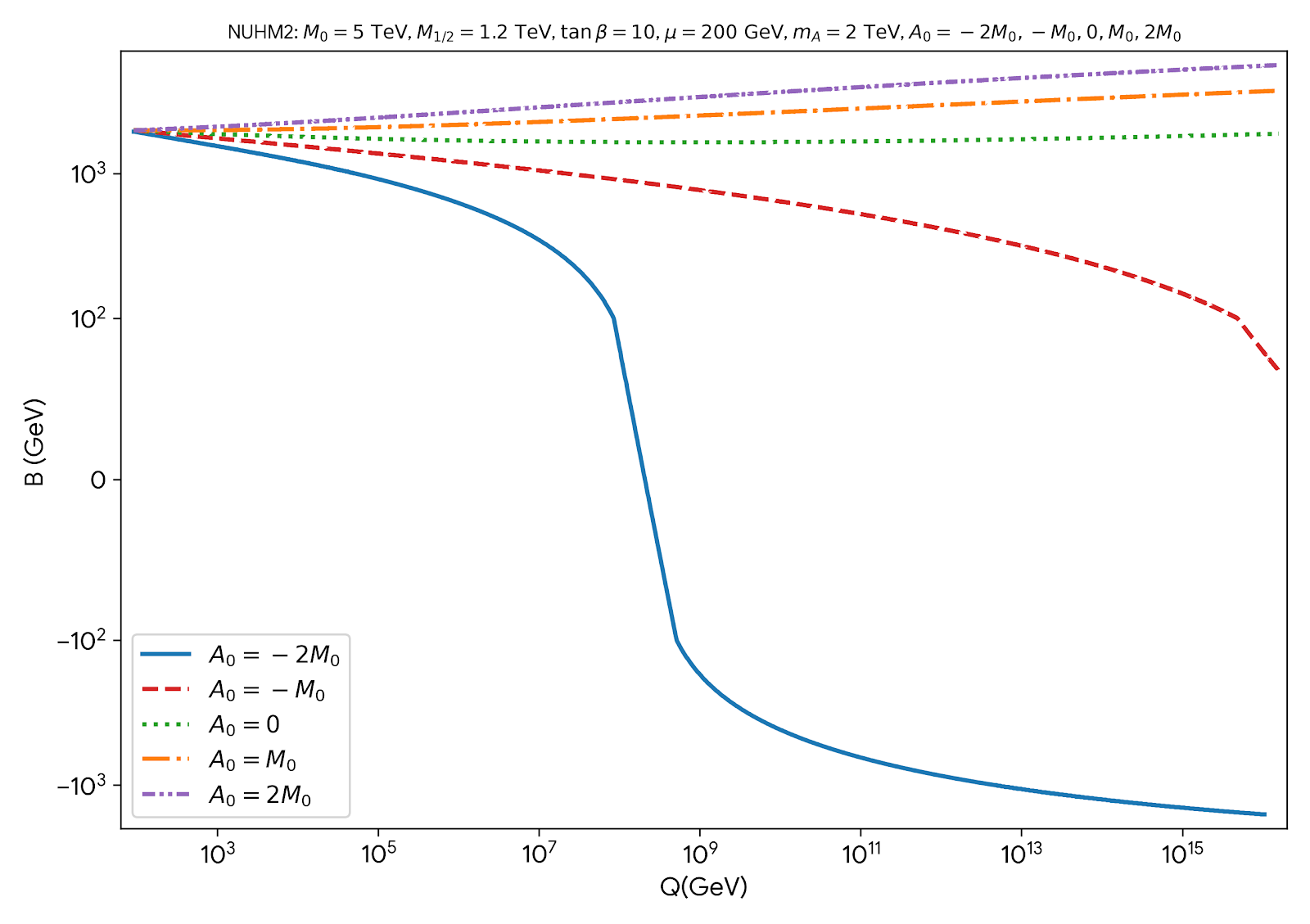}}
    \caption{Plot of $B$ vs. $Q$ evolution in the NUHM2 model with
      $m_0(1,2,3)=5$ TeV, $m_{1/2}=1.2$ TeV, $\tan\beta =10$ and $m_{H_d}=2$ TeV,
      with $m_{H_u}=\mu =200$ GeV but for different $A_0$ values.
            \label{fig:B}}
\end{figure}

The evolution of $B$ from $Q(weak)$ to $Q(m_{GUT})$ is shown for our BM
point in Fig. \ref{fig:B} but allowing for different values of $A_0$
ranging between $-2m_0$ to $+2m_0$.
For $A_0=+2m_0$, then the gauge contributions largely cancel the Yukawa
contributions in Eq. \ref{eq:dBdt}, and so the evolution of $B(Q)$ vs $Q$
is rather flat. 
For large negative $A_0$ values, then the gauge and Yukawa contributions reinforce leading to a large negative slope in Eq. \ref{eq:dBdt} and consequently large changes in $B(Q)$.
In fact, for very large negative $A_0$ values, then $B(Q)$ is required to
have negative values for large $Q\gg m_{weak}$.

\section{Living dangerously\cite{Arkani-Hamed:2005zuc}}
\label{sec:results}

\begin{figure}[htb!]
\centering
    {\includegraphics[height=0.4\textheight]{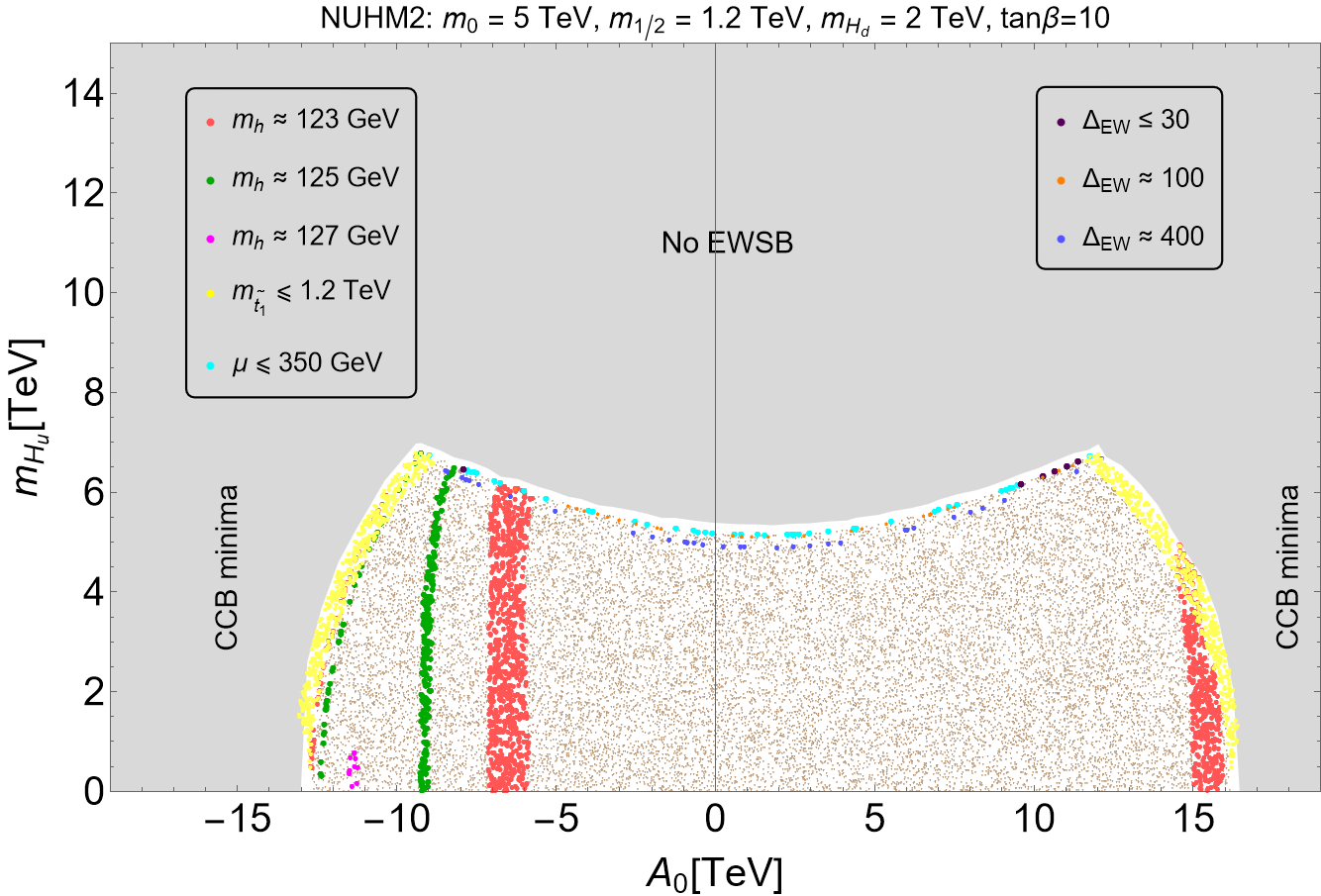}}\\
        {\includegraphics[height=0.45\textheight]{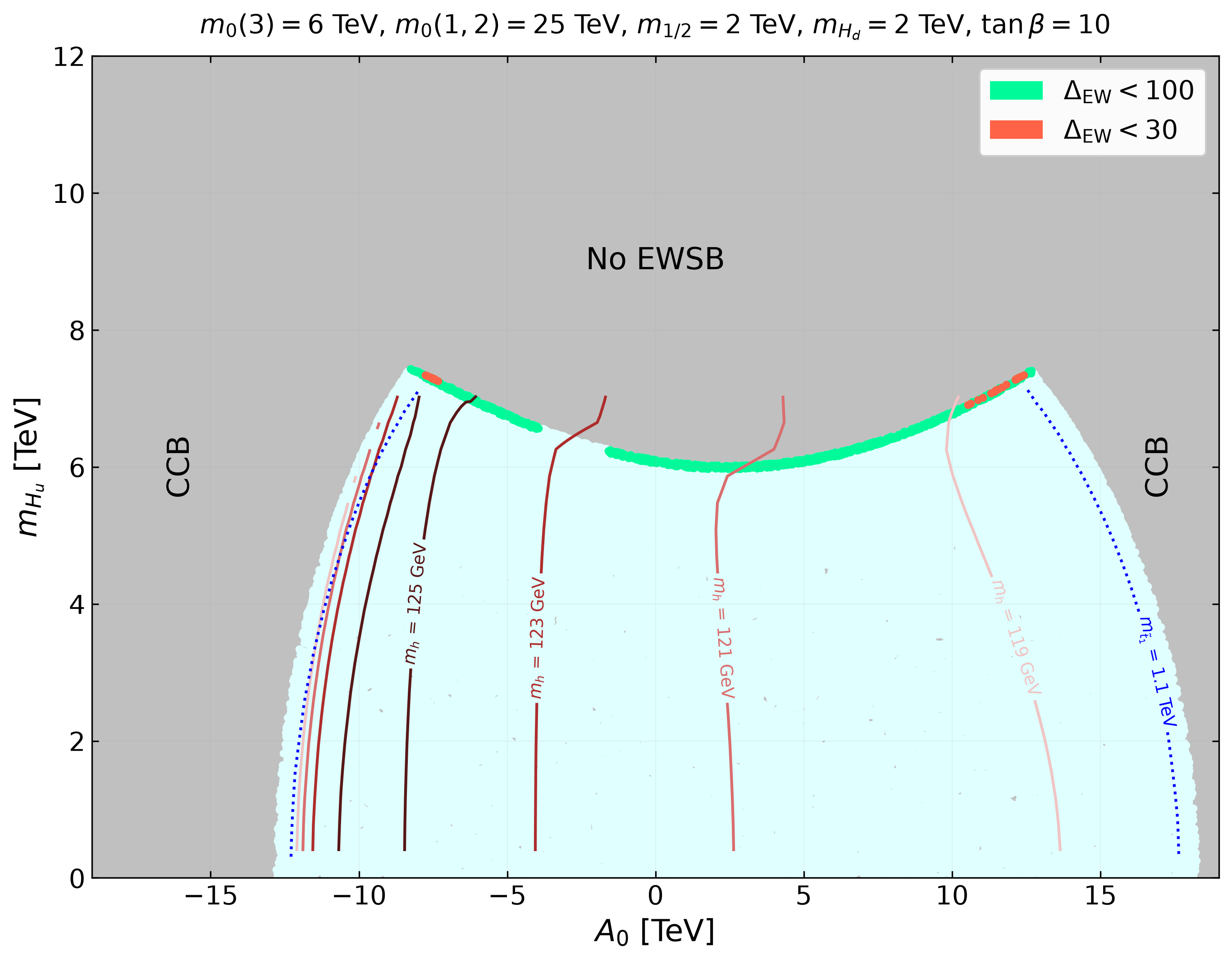}}
    \caption{Plot of {\it a}). $A_0$ vs. $m_{H_u}$ plane in NUHM2 model with
      $m_0(1,2,3)=5$ TeV, $\tan\beta =10$, $m_{1/2}=1.2$ TeV
      and $m_{H_d}=2$ TeV.
      In {\it b}), we make the same plot for NUHM3 with $m_0(1,2)=25$ TeV,
      $m_0(3)=6$ TeV, $m_{1/2}=2$ TeV and other parameters the same as for frame {\it a}).
            \label{fig:mHu_v_A}}
\end{figure}

To depict the theme of this paper, we show in Fig. \ref{fig:mHu_v_A}{\it a})
the $A_0$ vs. $m_{H_u}(Q=m_{GUT})$ plane in the NUHM2 model with
parameters $m_0=5$ TeV, $m_{1/2}=1.2$ TeV, $\tan\beta =10$ and $m_{H_d} =2$ TeV.
In frame {\it b}), we show the same plane but in the NUHM3 model with
$m_0(1,2)=25$ TeV which gives a decoupling/quasi-degeneracy solution to the SUSY
flavor and CP problems\cite{Baer:2019zfl,Baer:2024hpl}
(which are otherwise endemic to SUGRA-based SUSY-breaking models\cite{Soni:1983rm}).
(In the decoupling limit, then $m_{A,H,H^\pm}\sim m_{H_d}$).
Recall the high scale soft terms are preferred as large as possible on the
string landscape, so there is a statistical draw to large $|A_0|$ and also
to large $m_{H_u}$.
That translates to a draw to the upper-left or upper-right regions of
the parameter space that is shown.
But to attain an anthropic vacuum, one must stop short of the gray shaded
region where the upper-middle is labeled as {\it No EWSB} and the
left/right gray shaded sides have {\it CCB minima}.
The upper-arc just below the gray {\it No EWSB} region has small values of
$\Delta_{EW}$.
This is where $m_{H_u}(m_{GUT})$ is just large enough for EWSB to occur:
and if it gets much bigger, then $m_{H_u}^2$ doesn't get driven negative
and EWSB does not occur.
Also, the bigger $|A_0|$ gets, then the bigger $m_h$ becomes
(denoted by the red, green and magenta contours).
In the region where $|A_0|$ is close to the CCB-forbidden boundary,
the top-squark mass $m_{\tst_1}$ also
gets quite light and the $\Sigma_u^u(\tst_{1,2})$ terms decrease, making the
model more natural in the radiative correction sector.
Invoking maximal, but anthropic, values of both $m_{H_u}$ and $|A_0|$
puts us into the parameter space with natural SUSY where also $m_h\sim 125$ GeV: this illustrates how natural SUSY (as specified by low $\Delta_{EW}$)  is
{\it emergent}\cite{Baer:2022wxe,Baer:2024kms} from the string landscape due to the draw to large soft terms
of both $m_{H_u}$ and $|A_0|$.
Also, from this we may conclude that the LHC search for light higgsinos
and TeV-scale top-squarks are perhaps the most highly motivated
SUSY search channels.

\begin{figure}[htb!]
\centering
    {\includegraphics[height=0.5\textheight]{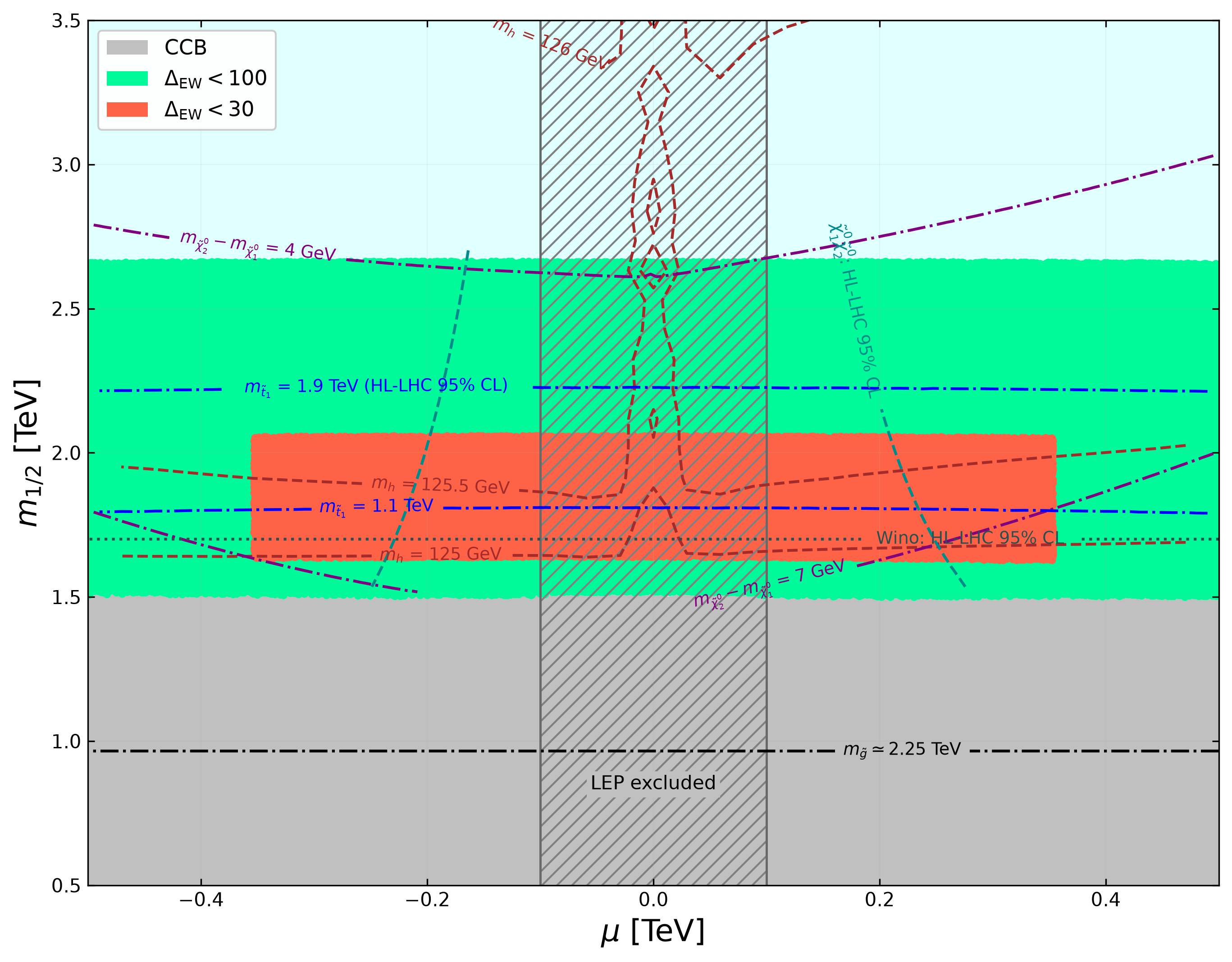}}
    \caption{Plot of $\mu$ vs. $m_{1/2}$ plane in the NUHM3 model
      with $m_0(1,2)=25$ TeV, $m_0(3)=6$ TeV, $\tan\beta =10$,
      $m_{A}=2$ TeV and $A_0=-1.2 m_0(3)$.
            \label{fig:mu_v_mhf}}
\end{figure}

In Fig. \ref{fig:mu_v_mhf}, we show the panoramic $\mu$ vs. $m_{1/2}$
plane for the NUHM3 model with $m_0(1,2)=25$ TeV (for the landscape-motivated
decoupling solution to the SUSY flavor/CP problems).
We also take $m_0(3)=6$ TeV, $\tan\beta =10$, $m_A=2$ TeV and $A_0=-1.2 m_0(3)$.
The rather large $m_0(1,2)$ value leads to CCB minima in the scalar potential
for the lower portion of $m_{1/2}$ values. In this region, we see the
LHC gluino mass limit contour of $m_{\tg}=2.25$ TeV.
This helps illustrate that 20th century SUSY searches in the region of
low $m_0$ and $m_{1/2}$ (as in the CMSSM/mSUGRA model) were searching in a
heavily landscape disfavored/disallowed region, and provides some rationale as
to why SUSY has not yet been discovered at LHC.
From the Figure, we also see the central LEP2 excluded region
$|\mu |\alt 100$ GeV where chargino pair production at LEP2 would have occurred.
The central red-shaded region denotes where $\Delta_{EW} <30$, {\it i.e.}
the natural SUSY region.
Relaxing $\Delta_{EW}<100$ leads to the green shaded region.
We also show contours of light Higgs mass $m_h$
(from FeynHiggs) which range from
$\sim 125$ GeV to 126 GeV (dashed curves at bottom/top of the plot).

Pertinent to LHC SUSY searches, we also show in Fig. \ref{fig:mu_v_mhf}
the calculated reach of high luminosity LHC (HL-LHC)
(with $\sqrt{s}=14$ TeV and 3000 fb$^{-1}$) for light higgsino pair
production $pp\to\tchi_1^0\tchi_2^0$ with $\tchi_2^0\to\tchi_1^0\ell^+\ell^-$:
the soft opposite-sign dilepton, jet plus $\eslt$ signature
(OSDLJMET)\cite{Baer:2021srt} as the sloping dashed contours
(for earlier work, see {\it e.g.} \cite{Baer:2011ec,Han:2014kaa,Baer:2014kya}).
The low $|\mu |$ region between these contours should be accessible to HL-LHC.
Some natural red region at higher $|\mu |$ values lies beyond HL-LHC reach.
Relevant to this signature, we also show contours of dilepton invariant mass
$m(\ell^+\ell^-)<m_{\tchi_2^0}-m_{\tchi_1^0}$ ranging from 4 GeV (lower $m_{1/2}$)
to 8 GeV (near top of green region).
This shows that the pure higgsino limit is not really attained in natural SUSY.
We show as the dotted contour around $m_{1/2}\sim 1.65$ TeV the HL-LHC reach for
wino pair production via an assortment of signatures including
same-sign dibosons (SSdBs)\cite{Baer:2013yha,Baer:2017gzf,Baer:2023olq,Baer:2025zqt}.

We also show contours of lighter top-squark mass $m_{\tst_1}=1.1-1.9$ TeV
(dot-dashed blue curves). The present reach of LHC from Run 2 data is around
1.1 TeV whilst the expected reach of HL-LHC is to $m_{\tst_1}\sim 1.9$
TeV\cite{Baer:2023uwo,Baer:2025zqt}.
While the HL-LHC can explore only part of the natural SUSY
region via the OSDLJMET signal, it can explore the entire region via
top-squark pair production $pp\to\tst_1\tst_1^*$.

\section{Conclusions}
\label{sec:conclude}

Large trilinear soft $A$ terms do two jobs at once in the MSSM.
The stop mixing term in Eq.~(\ref{eq:mh}) lifts $m_h$ to its measured value
only for weak scale $|A_t|$ of several TeV,
and the same large negative $A_t$ produces cancellations in the top-squark loop
contributions $\Sigma_u^u(\tilde t_{1,2})$ to the weak scale in
Eq.~(\ref{eq:Sigmauu}), lowering finetuning measure $\Delta_{EW}$.
The LHC Higgs mass therefore points to a SUSY breaking sector that supplies
$A\sim m_{3/2}\sim m_{weak}$.
That is the expectation of gravity mediation; gauge, anomaly and
gaugino mediation deliver tiny $A$-terms and must instead obtain
$m_h\simeq 125$ GeV with unnaturally heavy top squarks.
Within SUGRA, $A$-terms share the requirements of gaugino masses:
a gauge-singlet hidden sector field with an $F$-term vev and a broken
$R$-symmetry, which can be provided by metastable SUSY breaking.

On the string landscape the soft terms are drawn to large values by the
power-law distribution of Eq.~(\ref{eq:fSUSY}), and only the conditions
for a livable vacuum halt the draw: proper electroweak symmetry breaking
with no charge-or-color breaking minima, and a pocket-universe weak
scale inside the ABDS window of Eq.~(\ref{eq:abds}).
The $A$-terms enter both conditions through their renormalization group
evolution.
For the benchmark of Fig.~\ref{fig:Aterms}, $A_\tau$ hardly
runs while $A_t$ runs substantially, with $A_0\simeq-1.6\,m_0$ at $m_{GUT}$
arriving at $A_t(weak)\simeq-5$ TeV, the range needed for $m_h\simeq 125$ GeV.
At the same time $A_t^2$ feeds $X_t$ in the $m_{U_3}^2$ RGE of
Eq.~(\ref{eq:dmU3dt}), and for $A_0\simeq-2m_0$ or $+2.5m_0$ the
top-squark soft mass is driven tachyonic above the weak scale
(Fig.~\ref{fig:mU3}), producing
CCB vacua that must be vetoed.
Between these limits, $A_0=\pm m_0$ enhances the downward running
of $m_{U_3}^2$, giving lighter top squarks and a more natural spectrum.

The bilinear term is usually traded for $\tan\beta$ through the minimization
condition Eq.~(\ref{eq:B}), but its evolution carries information of its own.
Large $B(weak)$ requires $\tan\beta\to 1$ (Fig.~\ref{fig:BvTanb}),
which inflates the $m_{H_d}/\tan\beta$ contribution in Eq.~(\ref{eq:mzs})
and pushes $m_Z^{PU}$ out of the ABDS window;
large $B(weak)$ and large $m_{H_d}$ cannot coexist.
The slope of $B(Q)$ is set by the sign of $A_0$ in Eq.~(\ref{eq:dBdt}):
for $A_0=+2m_0$ the gauge and Yukawa terms nearly cancel and $B$ runs flat,
while for large negative $A_0$ they reinforce and $B(m_{GUT})$
is driven negative (Fig.~\ref{fig:B}).

The picture that emerges in the $A_0$ vs. $m_{H_u}$ plane
(Fig.~\ref{fig:mHu_v_A}) is the theme of this paper.
The statistical pull toward large $|A_0|$ and large $m_{H_u}$ carries the
parameters to the edge of the anthropically allowed region,
bounded above by the absence of EWSB and on either side by CCB minima.
The arc just inside the no-EWSB boundary is where $m_{H_u}^2$ is barely driven
negative and $\Delta_{EW}$ is smallest;
the strip adjoining the CCB boundary is where $m_h$ reaches
125 GeV and where $m_{\tilde t_1}$ becomes light enough that the
$\Sigma_u^u(\tilde t_{1,2})$ shrink.
Natural SUSY with the measured Higgs mass is then not an input but the
place where the landscape draw comes to rest, just short of catastrophic vacua.
The same reasoning in the $\mu$ vs. $m_{1/2}$ plane (Fig.~\ref{fig:mu_v_mhf})
shows the low-$m_0$, low-$m_{1/2}$ region explored by CMSSM-type searches
to be landscape-disfavored or CCB-excluded, which offers one rationale for the
absence of a SUSY signal to date.
These conclusions are drawn from specific NUHM2 and NUHM3
benchmarks, but the mechanisms behind them, the $A_t^2$ term in the $m_{U_3}^2$
RGE and the cancellations in Eq.~(\ref{eq:Sigmauu}),
do not depend on the benchmark choice.

Light higgsinos and TeV-scale top squarks are accordingly the most promising targets for future LHC SUSY searches.
In the NUHM3 plane of Fig.~\ref{fig:mu_v_mhf}, HL-LHC with 3000 fb$^{-1}$
reaches part of the $\Delta_{EW}<30$ region through the soft opposite-sign
dilepton plus jet plus $\MET$ signature from higgsino pair production,
with dilepton mass edges of $4-8$ GeV, and the same-sign diboson channel
from wino pairs probes $m_{1/2}\simeq 1.65$ TeV.
Top-squark pair production covers the entire natural region:
the present Run~2 limit is near $m_{\tilde t_1}\simeq 1.1$ TeV, while the expected
HL-LHC reach of $m_{\tilde t_1}\simeq 1.9$ TeV extends past the top of the
$\Delta_{EW}<30$ band.
The large $A$-terms that the landscape favors are thus also what place a
decisive test of this scenario within reach of the HL-LHC.

{\it Acknowledgements:} 

VB gratefully acknowledges support from the William F. Vilas estate.
HB and KZ gratefully acknowledges support from the Avenir Foundation.


\bibliography{A}
\bibliographystyle{elsarticle-num}

\end{document}